# A characterization of 2-player mechanisms for scheduling [*]

George Christodoulou[†]   Elias Koutsoupias[‡]   Angelina Vidali [§]

October 27, 2018

**Abstract**

We study the mechanism design problem of scheduling unrelated machines and we completely characterize the decisive truthful mechanisms for two players when the domain contains both positive and negative values. We show that the class of truthful mechanisms is very limited: A decisive truthful mechanism partitions the tasks into groups so that the tasks in each group are allocated independently of the other groups. Tasks in a group of size at least two are allocated by an affine minimizer and tasks in singleton groups by a task-independent mechanism. This characterization is about all truthful mechanisms, including those with unbounded approximation ratio.

A direct consequence of this approach is that the approximation ratio of mechanisms for two players is 2, even for two tasks. In fact, it follows that for two players, VCG is the unique algorithm with optimal approximation 2.

This characterization provides some support that any decisive truthful mechanism (for 3 or more players) partitions the tasks into groups some of which are allocated by affine minimizers, while the rest are allocated by a threshold mechanism (in which a task is allocated to a player when it is below a threshold value which depends only on the values of the other players). We also show here that the class of threshold mechanisms is identical to the class of additive mechanisms.

## 1 Introduction

Algorithmic mechanism design is an important area between computer science and economics. The two most fundamental problems in this area are the problem of scheduling unrelated machines [31] and the problem of combinatorial auctions [23, 13, 8]. Here we are dealing with the scheduling problem, but our main result which is the characterization of truthful mechanisms for two players extends naturally to the more general domain of combinatorial auctions. In the scheduling problem, there are $n$ players (machines) and $m$ tasks to be executed on these machines. Each task $j$ needs time $t_{ij}$ on machine $i$. We want to allocate the tasks to machines in a way that minimizes the makespan (the time required to finish all tasks). The problem is that the machines are selfish and will not reveal the true values (we assume that only machine $i$ knows the true values $t_{ij}$).

---

[*]Supported in part by IST-15964 (AEOLUS), IST-2008-215270 (FRONTS), and the Greek GSRT.
[†]Max-Planck-Institut für Informatik, Saarbrücken, Germany Email: gchristo@mpi-inf.mpg.de
[‡]Department of Informatics, University of Athens, Email: elias@di.uoa.gr
[§]Department of Informatics, University of Athens Email: avidali@di.uoa.gr

When we depart from the classical design of algorithms and try to extend it to mechanisms, we face the problem that these algorithms have to deal with selfish agents, who may not be truthful. This restricts the repertoire of available algorithms and brings forth the question of what kind of mechanisms are available in this framework.

A mechanism consists of two parts, the allocation algorithm and the payment functions, one for each player. Each player $i$ declares its own execution times $t_i$. The mechanism collects all the declarations $t$ and allocates the tasks according to an allocation function $a : R^{n \times m} \to \{1, \ldots, n\}^m$ from the set of all execution times to the set of partitions of $m$ tasks to $n$ players. It is more convenient to denote an allocation using the characteristic variables: $a_{ij}$ is an indicator variable for task $j$ to be allocated to machine $i$. The mechanism also pays each player $i$ a payment $p_i$. The payment depends on the declared values $t$ and indirectly on the allocation. A mechanism is truthful, if no player has incentive to lie. We are dealing here with the standard and more restricted notion of truthfulness, dominant truthfulness, in which a player has no incentive to lie for every value of the other players. It is well-known that in truthful mechanisms, the payment to player $i$ depends on the values $t_{-i}$ of the other players and on the allocation $a_i$ of player $i$: $p_i = p_i(a_i, t_{-i})$.

The allocation of the mechanism to player $i$ is given by the argmin expression $a_i = \mathrm{argmin}_a\{a_i \cdot t_i - p_i(a_i, t_{-i})\}$. The allocations to players must be consistent, i.e., every task is allocated to exactly one machine. The question is what type of allocation algorithms and payment schemes satisfy this property.

There is a simple answer to this question: A mechanism is truthful if and only if it satisfies the *Monotonicity Property*: If $a$ and $a'$ are the allocations of the mechanism for inputs $t$ and $t'$ which differ only on the values of player $i$, then we must have $\sum_{j=1}^{m}(a_{ij}-a'_{ij})(t_{ij}-t'_{ij}) \leq 0$. One nice property of this characterization is that it does not involve the payments at all. Since we usually care about the allocation part of mechanisms, this property focuses exactly on the interesting part. Unfortunately, although this is a necessary and sufficient condition [33], it is not very useful because it is a local and indirect property. The best way to clarify this point is to consider the case of mechanism design in unrestricted domains. In such domains, the same monotonicity property characterizes the truthful mechanisms. However, there is a much more direct characterization due to Roberts [19]: The class of truthful mechanisms for the unrestricted domain is very limited and contains exactly the class of affine maximizers. An important open problem is to come up with a similar characterization for the scheduling problem and combinatorial auctions. This work resolves this question for 2 players.

For the scheduling problem, very few mechanisms are known to be truthful. The principal example is the VCG mechanism [34, 12, 16] (or second-price mechanism) and its generalization, the affine minimizers [23]. The VCG mechanism allocates each task independently to the machine with minimum value, and pays the machine the second minimum value. VCG can be generalized in two ways and retain its truthfulness. The first generalization is the task-independent mechanisms, which allocate each task independently of the rest. The second generalization is the affine minimizers, which multiply the value of each player by some constant, but more importantly, they alter the value of each allocation by a constant. It is this set of additive constants, one per allocation, which make this generalization different than the first generalization.

Both these generalizations are known to be truthful, but they make very poor algorithms. The reason is that they allocate each task independently, or almost independently. The question is whether there are other truthful mechanisms. Unfortunately, we show here that this is not the case for 2 players (with a single uninteresting exception). For 3 or more players however, we know that there are some truthful mechanisms, which are slightly more general; we call them *threshold* mechanisms: For each task $j$ and each player $i$, there are thresholds $h_{ij}$ such that the player gets the task if and only if the value $t_{ij}$ is less than $h_{ij}$; the characterizing property of threshold mechanisms is that the threshold depends on the values of the other players, otherwise every mechanism can be expressed with thresholds (see Figure 1[$c_1 = 0$] for the geometric fingerprint of these mechanisms which partition the space with orthogonal hyperplanes). For two players the definition of threshold algorithms coincides with the definition of task-independent mechanisms. For 3 or more players, the class of threshold mechanisms is richer than the class of task-independent mechanisms (see for example [15, subsection 4.3]).

The question is whether, the affine maximizers and the threshold mechanisms exhaust the class of truthful mechanisms. The answer appears at first to be negative: For example, the mechanism that allocates all tasks to one player, the one with minimum sum of execution times, is truthful but it is neither affine minimizer nor threshold. However, this negative answer is not satisfactory because some allocations are never used, no matter how high or low are the values of the players. (One of the undesired properties of these mechanisms is that they have unbounded approximation ratio.) In contrast, we usually require that mechanisms have a much stronger property: decisiveness. A mechanism is called *decisive* when a player can enforce an outcome (allocation), by declaring very high or very low values. In fact, for the scheduling problem, it makes more sense to consider *locally-decisive* mechanisms: A player can enforce his allocation by declaring very low or high values, but cannot determine how the remaining tasks are allocated among the other players. When there are only two players, the notions of decisiveness and local-decisiveness coincide, but for 3 or more players decisiveness is a stronger property. We will restrict the discussion in this work to local-decisiveness.

A natural question is to characterize the decisive truthful algorithms. Unfortunately, by restricting our interest to decisive algorithms, we leave out important truthful specimens because some affine minimizers are not decisive: in some cases, a task will not be allocated to a player even when he declares 0 value for the task. To circumvent this problem, we allow negative values and we characterize the decisive truthful mechanisms for the domain of real values (both positive and negative). These algorithms include the affine minimizers and the monotone threshold algorithms; furthermore, every such algorithm is also truthful (but not necessarily decisive) for the nonnegative domain. By allowing negative values, we obtain not only a clean characterization but a useful one too, because we can still use it to argue about the approximation ratio for nonnegative values.

Our characterization leads us to conjecture:

**Conjecture 1.** For any number of players, a decisive truthful mechanism partitions the tasks into groups. Each group of tasks is allocated by either an affine minimizer or a threshold mechanism.

In this work, we show that the conjecture holds for 2 players (Theorem 1). If this turns out to be true for more players, it will show that the class of truthful mechanisms is limited to a few algorithms with poor performance. This will also apply directly to richer domains, such as combinatorial auctions (the richer the domain the more restrictive the class of truthful mechanisms). In fact, for 2 players our theorem is stronger than the conjecture in two aspects:

first, in the statement of the conjecture, we can replace the threshold mechanisms with the stronger notion of task-independent mechanisms; this is possible because for 2 players the thresholds mechanisms are also task-independent. Second, for 2 players, each part is allocated independently of the rest. This is not necessarily true for 3 or more players, as the following example shows:

*Example* 1. Consider a mechanism with 3 players and 3 tasks, where the first 2 tasks are allocated by an affine minimizer, while the third task is allocated by a threshold mechanism as follows: the task is given to the first player when he has minimum value ($t_{31} \leq \min\{t_{23}, t_{33}\}$); otherwise it is given to the second player if and only if $t_{23} + t_{11} \leq t_{33}$. Notice that the value $t_{11}$ of the affine minimizer affects the part of threshold mechanisms.

On a side note, when the affine minimizer in the above mechanism is the VCG mechanism, we obtain an example of a truthful threshold mechanism which is not task-independent.

The decisiveness restriction is necessary to keep our presentation simple. We don't know whether one could drop it and still preserve the essence of the proof. In fact, for the case of two tasks we only assume decisiveness for 3 allocations (in the sense that a mechanism is decisive for an allocation if each one of the players can impose this allocation by changing his values while the values of the other player remain fixed).

In our presentation we deal a lot with payments and, since we are only interested in the difference of payments, we will use the following notation

$$f^i_{a:a'}(t_{-i}) = p_i(a'_i, t_{-i}) - p_i(a_i, t_{-i}).$$

For simplicity, we write $f_{a:a'}$ in place of $f^1_{a:a'}$. We also represent the allocations using only the allocation of player 1, since the allocation of player 2 can be inferred. For example, we write $f_{00:10}$ for the difference in payments of player 1 when he gets only task 1 and when he gets no task. There is an extra reason to define $f_{a:a'}$: at some point in our proof, we will use the inverse function $f^{-1}_{a:a'}$.

The main reason for using negative values in our characterization is that the values $f_{a:a'}$, being the differences of payments, can take negative values.

As we mentioned, the allocation of a mechanism can be expressed with argmin expressions, one for every player: $a_i = \operatorname{argmin}_a\{a_i \cdot t_i - p_i(a_i, t_{-i})\}$. For two players and two tasks, we essentially seek the payments that satisfy the following equation, which expresses the fact that the allocations for the two players must be consistent (i.e. each task is allocated exactly once):

$$\operatorname{argmin}\{t_{11} + t_{12} - p_1(11, t_2), t_{11} - p_1(10, t_2), t_{12} - p_1(01, t_2), -p_1(00, t_2)\} =$$
$$\operatorname{argmin}\{-p_2(11, t_1), t_{22} - p_2(10, t_1), t_{21} - p_2(01, t_1), t_{21} + t_{22} - p_2(00, t_1)\}.$$

Therefore the problem of characterizing the argmin mechanisms for two players and two tasks boils down to the following simple question: Which payments $p$ satisfy the above equation? This is precisely the problem that we are trying to solve here.

The following theorem provides the answer, which applies also to any number of tasks. But first we give a precise definition of the affine minimizers:

**Definition 1** (Affine minimizers)**.** A mechanism is an affine minimizer if there are constants $\lambda_i > 0$ (one for each player $i$) and $\gamma_a$ (one for each of the $n^m$ allocations) such that the mechanism selects the allocation $a$ which minimizes $\sum_i \lambda_i a_i t_i + \gamma_a$.

We now state our main result:

**Theorem 1.** *For the scheduling problem with real values every decisive truthful mechanism for 2 players partitions the tasks into groups so that the tasks in each group are allocated independently of the other groups. Tasks in a group of size at least two are allocated by an affine minimizer and tasks in singleton groups by a task-independent mechanism.*

## 2  Related Work

The scheduling problem on unrelated machines is one of the most fundamental problems in combinatorial optimization [18]. Lenstra, Shmoys, and Tardos [25] gave a 2-approximation polynomial-time algorithm for the classical version of the problem. They also showed that the problem cannot be approximated in polynomial time within a factor less than 3/2.

Here we study its mechanism design version which was introduced by Nisan and Ronen in their paper [31, 32] that initiated the algorithmic theory of Mechanism Design. They gave a truthful $n$-approximate (polynomial-time) algorithm (where $n$ is the number of machines); they also showed that no mechanism (polynomial-time or not) can achieve approximation ratio better than 2 when there are at least three tasks. We strengthen this result by proving that it holds even for only two tasks.

The lower bound for deterministic mechanisms has been improved in [11] to 2.41 (this is the best-known lower bound for 3 machines) and [20] to 2.618 for $n \to \infty$ machines.

Nisan and Ronen [31] also gave a randomized truthful mechanism for two players, that achieves an approximation ratio of 7/4. Mu'alem and Schapira [29] proved a lower bound of $2 - \frac{1}{n}$ for any randomized truthful mechanism for $n$ machines and generalized the mechanism in [31] to give a $7n/8$ upper bound. Recently Lu and Yu [27, 26] give an 1.59-approximation universally truthful randomized algorithm.

In another direction, [10] shows that no fractional truthful mechanism can achieve an approximation ratio better than $2 - 1/n$. It also shows that fractional algorithms that treat each task independently cannot do better than $(n+1)/2$ and this bound is tight.

Lavi and Swamy [24] consider a special case of the same problem—namely when the processing times have only two possible values low or high—and devise a deterministic 2-approximation truthful mechanism. Recently Yu [35] extends their results.

A special case of the problem is the problem on related machines in which there is a single value (instead of a vector) for every machine, its speed. Myerson [30] gave a characterization of truthful algorithms for this kind of problems (one-parameter problems), in terms of a monotonicity condition. Archer and Tardos [4] found a similar characterization and using it obtained a variant of the optimal algorithm which is truthful (albeit exponential-time). They also gave a polynomial-time randomized 3-approximation mechanism, which was later improved to a 2-approximation, in [2]. This mechanism is truthful in expectation. Auletta et al. [6] gave a 4-approximation deterministic algorithm for any fixed number of machines. Andelman, Azar, and Sorani [1] gave a 5-approximation deterministic truthful mechanism, for any number of machines. Kovacs improved the approximation ratio to 3 [21] and to 2.8 [22].

Much more work has been done in the context of combinatorial auctions (see for example [3, 8, 9, 13, 7, 14] and the references within).

Our approach of aiming at a complete characterization of truthful mechanisms, regardless of approximation ratio, is analogous to Roberts [19] result for unrestricted domains, but also resembles the approach in [23, 5], and it was influenced by the limitations of the current methods in establishing the known lower bounds [31, 11, 20].

Saks and Yu [33] proved that, for mechanism design problems with convex domains, which includes the scheduling problem, a simple necessary monotonicity property, between different inputs (and without any reference to payments) is also sufficient for truthful mechanisms, generalizing results of [17, 23]. Monderer [28] shows that the domain cannot be further generalized in the case of quasi-linear utility functions, because in this case a domain of valuations is a monotonicity domain iff its closure is convex. Our characterization of additive mechanisms without any reference to payments is in the same direction with this work.

A very recent paper [15] by Dobzinski and Sundararajan is very close in spirit to this work. Dobzinski and Sundararajan restrict their attention to mechanisms with bounded approximation ratio. They show that the truthful mechanisms with bounded approximation ratio are task-independent. In contrast, our work provides a more complete characterization of all mechanisms including those with unbounded approximation ratio.

## 3 The characterization of decisive mechanisms for 2 tasks

Our main theorem is based on the following theorem which applies to 2 players and 2 tasks and which is the subject of this section.

**Theorem 2.** *For the scheduling problem with real values the decisive truthful mechanisms for 2 players and 2 tasks are either task-independent or affine minimizers. The same characterization applies to mechanisms that are decisive for only three outcomes.*

We proceed in our proof carefully, revealing gradually the properties of $f_{a:a'}$. We assume here that the payments take real (positive or negative) values, so that $f_{a:a'}$ is also a real function. An indispensable part of the proof is the following lemma.

**Lemma 1.** *For allocations $a$ and $a'$ that differ in only one task, the quantity $f_{a:a'}(t_2)$ depends only on $(a - a') \cdot t_2$ (and therefore it depends on only one variable).*

*Proof.* This lemma holds for every number of tasks. We will first prove the lemma for $m = 2$ tasks. We will focus on the case of $a = 00$ and $a' = 10$ since the other cases are very similar.

We will show by contradiction that $f_{00:10}(t_{21}, t_{22})$ does not depend on $t_{22}$. Suppose that there are $t_{21}$, $t_{22}$, and $t'_{22}$ with $t_{22} \neq t_{22'}$ with $f_{00:10}(t_{21}, t_{22}) < f_{00:10}(t_{21}, t'_{22})$.

From the definition of $f_{00:10}(t_{21}, t_{22})$, the tasks of the form

$$\begin{pmatrix} f_{00:10}(t_{21}, t_{22}) + \epsilon & \infty \\ t_{21} \star & t_{22} \star \end{pmatrix}$$

have the indicated allocation for every $\epsilon > 0$, *where $infty$ indicates an arbitrarily high value which guarantees that the second task will not be allocated to player 1 (i.e., $\infty$ is greater than $\max\{f_{00:01}(t_2), f_{00:11}(t_2)\}$).*

Similarly, the tasks of the form

$$\begin{pmatrix} f_{00:10}(t_{21}, t'_{22}) - \epsilon \star & \infty \\ t_{21} & t'_{22} \star \end{pmatrix}$$

have the indicated allocation for every $\epsilon > 0$. As we mentioned before, $\infty$ denotes an arbitrarily high value. We assume of course that the two occurrences of this symbol above denote the same value.

By the Monotonicity Property, if we decrease the values of $t_{22}$ to $t'_{22}$ to $\min\{t_{22}, t'_{22}\}$, the allocations remain the same.

This leads to a contradiction when $\epsilon = (f_{00:10}(t_{21}, t'_{22}) - f_{00:01}(t_{21}, t_{22}))/2$, because the task

$$\begin{pmatrix} \frac{f_{00:10}(t_{21}, t_{22}) + f_{00:10}(t_{21}, t'_{22})}{2} & \infty \\ t_{21} & \min\{t_{22}, t'_{22}\} \star \end{pmatrix}$$

would have two allocations.

The proof can be extended to the case of $m > 2$ tasks: We reduce it to the $m = 2$ case by fixing all tasks except of two. For example, for every $t_2 = (t_{21}, t_{22}, t_{23})$ and $t'_2 = (t_{21}, t'_{22}, t'_{23})$ we have: $f_{000:100}(t_{21}, t_{22}, t_{23}) = f_{000:100}(t_{21}, t'_{22}, t_{23}) = f_{000:100}(t_{21}, t'_{22}, t'_{23})$. □

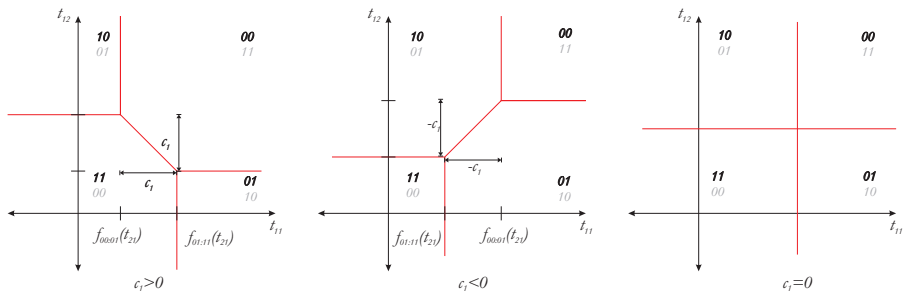

Figure 1: There are three ways a truthful mechanism can partition the input space of player 1 for fixed $t_2$, according to the sign of $c_1$. For $c_1 = 0$ you can see that there is a threshold $h_j(t_{2j})$ for each task $j$.

**Corollary 1.** *The quantities $c_1 = f_{01:11}(t_2) - f_{00:10}(t_2)$ and $c_2 = f^2_{10:00}(t_2) - f^2_{11:01}(t_2)$ do not depend on $t_2$.*

*Proof.* First observe that the following equality

$$f_{01:11}(t_2) - f_{00:10}(t_2) = f_{10:11}(t_2) - f_{00:01}(t_2),$$

follows directly from the definitions because both parts are equal to $p_1(11, t_2) - p_1(01, t_2) - p_1(10, t_2) + p_1(00, t_2)$. The above lemma states that $f_{01:11}(t_2)$ and $f_{00:10}(t_2)$ depend only on $t_{21}$. Consequently the left part of the above equality depends only on $t_{21}$. Similarly the right part of the above equality depends only on $t_{22}$. Therefore, both differences are constant (independent of $t_2$). We denote this constant by $c_1$ (the 1 stands for player 1), and we define, in a similar way, a constant $c_2$ for player 2. □

We can now define the regions of truthful mechanisms. For fixed $t_2$, let $R_{11}$ denote the set of values $t_1$ for which the mechanism allocates both tasks to player 1. Region $R_{11}$ which is defined by the following constraints:

$$t_{11} < f_{10:11}(t_{21})$$
$$t_{12} < f_{01:11}(t_{22})$$
$$t_{11} + t_{12} < f_{01:11}(t_{21}) + f_{00:01}(t_{22}).$$

Similarly, the inequalities for region $R_{00}$ are

$$t_{11} > f_{00:10}(t_{21})$$
$$t_{12} > f_{00:01}(t_{22})$$
$$t_{11} + t_{12} > f_{01:11}(t_{21}) + f_{00:01}(t_{22}).$$

There are similar constraints that define the other two regions $R_{10}$ and $R_{01}$. What happens at the boundaries, where the inequality becomes an equality is not determined by the Monotonicity Property. These undetermined values are a major source of difficulty in the characterization of the mechanisms.

From the above inequalities we get that the boundary between regions $R_{00}$ and $R_{11}$, if it exists, is of the form $t_{11} + t_{12} = f_{01:11}(t_{21}) + f_{00:10}(t_{22})$. Since a similar constraint holds for player 2 (in which the sum $t_{21} + t_{22}$ appears), one could be tempted to conclude that the boundary between allocations 00 and 11 is of the form $t_{11} + t_{12} = h(t_{21} + t_{22})$ for some function $h$. *Although this conclusion is exactly the one that we will eventually reach, the above argument is fallacious*: There is no justification to identify the boundary between regions $R_{00}$ and $R_{11}$ for the first player when $t_2$ is fixed and the boundary between the same regions for the second player when $t_1$ is fixed. In fact, we don't even know that the boundary is some surface when we consider the 4-dimensional space of $t$. We tried many shortcuts in our proof but we couldn't make them rigorous. This in part is reflected in the writing of the proof, in which we proceed carefully and use elementary and straightforward arguments.

To proceed to the characterization of mechanisms, we need to understand the functions $f_{00:10}$ and $f_{00:01}$. The first property is easy:

**Lemma 2.** *The functions $f_{01:11}$ and $f_{00:01}$ are nondecreasing.*

*Proof.* By symmetry, it suffices to establish the lemma only for the function $f_{00:01}$. Suppose that it is decreasing. Then there are $t_{21}, t'_{21}$ with $t_{21} < t'_{21}$ and $f_{00:01}(t_{21}) > f_{00:01}(t'_{21})$. Consider the instance

$$\begin{pmatrix} \frac{f_{00:01}(t_{21})+f_{00:01}(t'_{21})}{2} \star & 0 \star \\ t_{21} & \infty \end{pmatrix},$$

which should have the indicated allocation because $\frac{f_{00:01}(t_{21})+f_{00:01}(t'_{21})}{2} < f_{00:01}(t_{21})$. Similarly, the instance

$$\begin{pmatrix} \frac{f_{00:01}(t_{21})+f_{00:01}(t'_{21})}{2} & 0 \star \\ t'_{21} \star & \infty \end{pmatrix},$$

should have the indicated allocation. But since $t_{21} < t'_{21}$, this contradicts the Monotonicity Property. □

For most reasonable mechanisms, a stronger statement seems to apply for these two functions: that they are strictly increasing. This however is not generally true as the following example shows.

*Example* 2 (Task-independent mechanism but not-strictly increasing). Consider the task-independent mechanism with

$$f_{01:11}(t_{21}) = \begin{cases} t_{21} & t_{21} \leq 1 \\ 1 & 1 \leq t_{21} \leq 2 \\ t_{21} - 1 & 2 \leq t_{21} \end{cases}$$

and $f_{00:01}(t_{22}) = t_{22}$. The interesting property of this mechanism is that the function $f_{01:11}(t_{21})$ is not strictly increasing.

But we can show that the functions $f_{01:11}$ and $f_{00:01}$ are indeed strictly increasing when $c_1 \neq 0$. In fact, we show in the next lemma that either the functions are strictly increasing or they are like the following mechanism, which is not a decisive mechanism.

*Example* 3 (Mechanism with some oblivious player). Consider the mechanism with $f_{00:10}(t_{21}) = b_1$, $f_{00:01}(t_{22}) = b_2$ where $b_1$, $b_2$, and $c_1$ are constants. In this mechanism the first player decides independently of the values of the second player. For given values $t_1$ of the first player, the second player has the same allocation for every $t_2$. This mechanism is not decisive, since the second player cannot force all allocations.

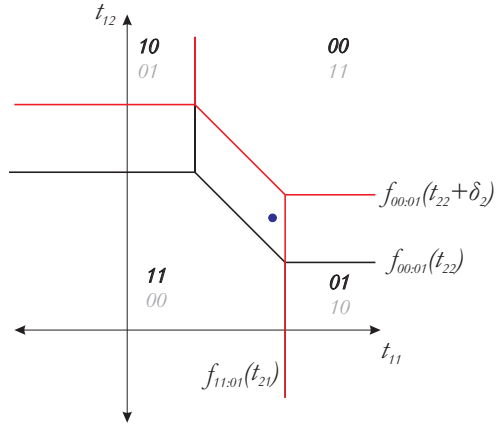

Figure 2: The points we use to prove Lemma 3

**Lemma 3.** *In a truthful mechanism with $c_1 \neq 0$ the functions $f_{01:11}$ and $f_{00:01}$ are either both strictly increasing or both constant. (The same holds for the pair $f_{00:10}$ and $f_{10:11}$.)*

*Proof.* We will prove the lemma for $c_1 > 0$ since the case $c_1 < 0$ is very similar. We will show that there are no $t_{21}$, $t_{22}$, $0 < \delta_2 < \delta_1$ such that

$$f_{01:11}(t_{21}) = f_{01:11}(t_{21} + \delta_1)$$
$$f_{00:01}(t_{22}) < f_{00:01}(t_{22} + \delta_2)$$

Before we prove this, we show that it implies the lemma. Indeed, if some of the functions $f_{01:11}$ and $f_{00:01}$ is not strictly increasing, say the function $f_{11:01}$, then it is somewhere constant, as we have already established in Lemma 2 that it is non-decreasing. Therefore there are $t_{21}$ and $\delta_1 > 0$ with $f_{01:11}(t_{21}) = f_{01:11}(t_{21} + \delta_1)$. But then for every $\delta_2 < \delta_1$, we must have $f_{00:01}(t_{22}) = f_{00:01}(t_{22} + \delta_2)$. It follows that $f_{00:01}$ is constant. This in turn (with similar reasoning) implies that $f_{00:10}$ is also constant.

We now return to the proof of the above statement. Towards a contradiction we assume that there is such a mechanism with $c_1 > 0$. For $0 < \epsilon < c_1/2$ we consider the inputs (see Figure 2)

$$\begin{pmatrix} f_{01:11}(t_{21}) - \epsilon \star & \frac{f_{00:01}(t_{22}) + f_{00:01}(t_{22} + \delta_2)}{2} + \epsilon \star \\ t_{21} & t_{22} + \delta_2 \end{pmatrix}$$

and

$$\begin{pmatrix} f_{01:11}(t_{21}) - \epsilon & \frac{f_{00:01}(t_{22}) + f_{00:01}(t_{22} + \delta_2)}{2} + \epsilon \\ t_{21} + \delta_1 \star & t_{22} \star \end{pmatrix}$$

The claim is that the mechanism will allocate the tasks as indicated by the stars, i. e., both tasks to the first player in the first input and both tasks to the second player in the second input. Indeed, it is easy to verify that the first input satisfies the inequalities that define $R_{11}$ and the second input satisfies the inequalities that define $R_{00}$.

But these allocations contradict the Monotonicity Property for player 2. The inputs are identical for the first player while for the second player the sum of the values are $t_{21} + t_{22} + \delta_2$ and $t_{21} + t_{22} + \delta_1$. Since we assumed that $\delta_2 < \delta_1$, the sum of the values of the second player in the first input is less than the sum in the second input. The allocations clearly violate the Monotonicity Property. □

The above lemma establishes that the mechanisms with $c_1 \neq 0$ are either one of the mechanisms of the Example 3 or both functions $f_{01:11}$ and $f_{00:01}$ are strictly increasing. As we consider decisive mechanisms, from now on we will consider only strictly increasing functions.

**Lemma 4.** *If $c_2 \neq 0$ then the functions $f_{01:11}$ and $f_{00:01}$ are bijections from $\mathbb{R}$ to $\mathbb{R}$.*

*Proof.* (Recall that the superscript 2 in $f^2_{00:10}$ stands for the second player. By definition, the allocations of the subscript however are still allocations of the first player. The corresponding allocations of player 2 can be obtained by changing the roles of 0 and 1.) We want to establish that the functions $f_{00:10}, f^2_{10:00}$ are inverse. We use the assumption $c_2 \neq 0$ only to guarantee that $f^2_{10:00}$ is strictly increasing.

From the definitions of the function, we get the following implications:

$$t_{11} < f_{00:10}(t_{21}) \quad \Rightarrow \quad f^2_{10:00}(t_{11}) \leq t_{21}$$
$$t_{11} > f_{00:10}(t_{21}) \quad \Rightarrow \quad f^2_{10:00}(t_{11}) \geq t_{21}$$

The claim is that the above conditions imply that the two functions are inverse.

Assume towards a contradiction that for some $t_{11}$ we have $f_{00:10}(f^2_{10:00}(t_{11})) = t'_{11}$ with $t'_{11} > t_{11}$ (the other case, $t'_{11} < t_{11}$, is similar). Then $(t_{11} + t'_{11})/2 > t_{11}$, which by the strictly increasing property of $f^2_{10:00}$ implies that $f^2_{10:00}((t_{11} + t'_{11})/2) > f_{00:10}(t_{11})$. On the other hand, $(t_{11} + t'_{11})/2 < t'_{11} = f_{00:10}(f^2_{10:00}(t_{11}))$ which by the above implications gives $f^2_{01:11}((t_{11} + t'_{11})/2) \leq f^2_{01:11}(t_{11})$, a contradiction. □

The assumption $c_2 \neq 0$ is essential in the above lemma. When $c_2 = 0$, there are mechanisms in which $f_{00:10}$ and $f_{00:01}$ are not bijections; for example, the mechanism of the following example.

*Example* 4 (Non-decisive task-independent mechanism). Consider the mechanism with $f_{00:10}(t_{21}) = e^{t_{21}}$, $f_{00:01}(t_{22}) = e^{t_{22}}$, and $c_1 = c_2 = 0$. This is a task-independent mechanism which is not decisive; the allocation cannot be expressed as argmin for both players.

**Lemma 5.** *The constants $c_1$ and $c_2$ are either both positive, both negative, or both 0.*

*Proof.* It suffices to show that if $c_1 > 0$ then it follows that $c_2 > 0$. Consider the tasks

$$\begin{pmatrix} f_{01:11}(t_{21}) - \epsilon \star & f_{00:01}(t_{22}) + 2\epsilon/3 \star \\ t_{21} & t_{22} \end{pmatrix} \quad \begin{pmatrix} f_{01:11}(t_{21}) & f_{00:01}(t_{22}) + \epsilon/3 \\ t_{21} \star & t_{22} \star \end{pmatrix}$$

It is straightforward to check the indicated allocations (for $c_1 > \epsilon > 0$).

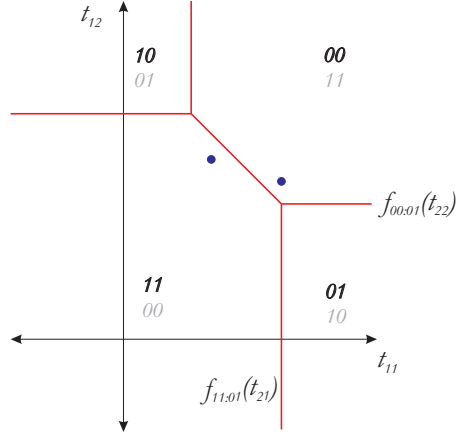

Figure 3: The points we take for Lemma 5

Let's denote the above values as: $t_{12} = f_{00:01}(t_{22}) + 2\epsilon/3$ and $t'_{12} = f_{00:01}(t_{22}) + \epsilon/3$. Now, if $c_2 \leq 0$, we should have that $t_{22} \geq f^2_{11:10}(t_{12})$ and $t_{22} \leq f^2_{11:10}(t'_{12}) + c_2$. (Consider the situation player 2 faces when the values of player 1 are fixed to $t_1$ and $t'_1$.) But since $t_{12} > t'_{12}$ and since $f^2_{11:10}$ is strictly increasing (as the inverse of a strictly increasing function) this leads to a contradiction when $c_2 \leq 0$. □

We now strengthen the characterization of $f_{00:10}$ and $f_{00:01}$ in the case when $c_1 \neq 0$.

**Lemma 6.** *For $c_1 \neq 0$, the functions $f_{00:10}$ and $f_{00:01}$ are semiperiodic and in particular they satisfy*

$$f_{00:10}(t_{21} + c_2) = f_{00:10}(t_{21}) + c_1$$

*and*

$$f_{00:01}(t_{22} + c_2) = f_{00:01}(t_{22}) + c_1.$$

*Proof.* Again by symmetry we need only to establish the lemma for $f_{00:10}$.

Notice first that $f_{01:11}$ is a bijection, for the same reasons that $f_{00:10}$ is a bijection. We also know that

$$f_{01:11}(t_{21}) = f_{00:10}(t_{21}) + c_1.$$

The associated equation for player 2 is

$$f_{00:10}^{-1}(t_{11}) = f_{01:11}^{-1}(t_{11}) + c_2.$$

We therefore have

$$\begin{aligned} f_{00:10}(t_{21}) + c_1 &= f_{00:10}(f_{00:10}^{-1}(f_{00:10}(t_{21}) + c_1)) \\ &= f_{00:10}(f_{01:11}^{-1}(f_{00:10}(t_{21}) + c_1) + c_2) \\ &= f_{00:10}(f_{01:11}^{-1}(f_{01:11}(t_{21})) + c_2) \\ &= f_{00:10}(t_{21} + c_2) \end{aligned}$$

The first equality is based on the trivial fact that $t_{11} = f_{00:10}(f_{00:10}^{-1}(t_{11}))$. Similarly for the last equality. The second and third equalities follow from the above mentioned equalities for player 2 and player 1. □

We will focus on the case of $c_1 > 0$ as the case $c_1 < 0$ is very similar. Consider the diagonal boundary between the regions $R_{11}$ and $R_{00}$. This boundary is on the line $t_{11} + t_{12} = f_{01:11}(t_{21}) + f_{00:01}(t_{22})$. We have $f_{00:11}(t_{21}, t_{22}) = f_{01:11}(t_{21}) + f_{00:01}(t_{22})$. The heart of the characterization is that the function $f_{00:11}(t_{21}, t_{22})$ depends only on the sum of $t_{21} + t_{22}$.

**Lemma 7.** *The function $f_{00:11}(t_{21}, t_{22}) = f_{01:11}(t_{21}) + f_{00:01}(t_{22})$ depends only on $t_{21} + t_{22}$, i. e., there is some function $h$ such that $f_{00:11}(t_{21}, t_{22}) = h(t_{21} + t_{22})$.*

*Proof.* Suppose not. That is suppose that there are $t_2$ and $t_2'$ such that $t_{21} + t_{22} = t_{21}' + t_{22}'$ and yet $f_{00:11}(t_{21}, t_{22}) < f_{00:11}(t_{21}', t_{22}')$. If the values differ, they have to differ for some $t_{21}$ and $t_{21}'$ that are very close.

Without loss of generality then we assume that $t_{21} < t_{21}' < t_{21} + c_2$.

This implies that $t_{22}' < t_{22} < t_{22'} + c_2$ and therefore

$$f_{00:01}(t_{22}) < f_{00:01}(t_{22}' + c_2) = f_{00:01}(t_{22}') + c_1.$$

Let $\epsilon$ be a positive parameter with $\epsilon < f_{00:11}(t_{21}', t_{22}') - f_{00:11}(t_{21}, t_{22})$ and $\epsilon < f_{01:11}(t_{22}') - f_{00:01}(t_{22})$. By the above inequalities, $\epsilon$ belongs to an open interval and more specifically it can take at least two distinct values. Consider then the values

$$\begin{aligned} t_{11} &= f_{01:11}(t_{21}) \\ t_{12} &= f_{00:01}(t_{22}) + \epsilon \end{aligned}$$

We can easily verify that the following inputs satisfy the boundary constraints of the appropriate regions ($R_{00}$ and $R_{11}$) and have the indicated allocations:

$$\begin{pmatrix} t_{11} & t_{12} \\ t_{21}\star & t_{22}\star \end{pmatrix} \quad \begin{pmatrix} t_{11}\star & t_{12}\star \\ t_{21}' & t_{22}' \end{pmatrix}$$

This means that, when we fix $t_1$, the points $t_2$ and $t_2'$ are on the boundary between regions $R_{11}$ and $R_{00}$ of player 2. Equivalently, that

$$t_{21} + t_{22} = f_{01:11}^{-1}(t_{11}) + f_{00:01}^{-1}(t_{12} - \epsilon).$$

(A similar equation holds for $t_2'$ which however is not different since we assumed that $t_{21} + t_{22} = t_{21}' + t_{22}'$). This equality should hold for every $\epsilon$ in some open interval. But this contradicts the fact that $f_{00:01}^{-1}$ is strictly increasing. □

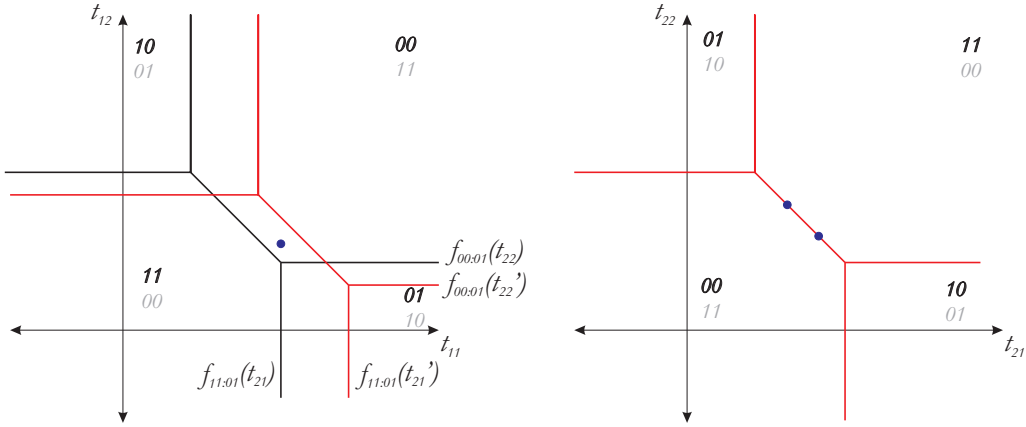

Figure 4: The point we use to prove Lemma 7 for player 1 and for player 2.

From the last lemma, we get that $h(t_{21} + t_{22}) = f_{01:11}(t_{21}) + f_{00:01}(t_{22})$. We claim that the functions involved are affine as the following lemma shows.

**Lemma 8.** *If for some real functions $h$, $h_1$, $h_2$ which are continuous at some point, we have $h(x+y) = h_1(x) + h_2(y)$, then all three functions are affine, i. e., they are of the form $ax+b$ for some constants $a$ and $b$.*

*Proof.* Let $g(x) = h(x) - h_1(0) - h_2(0)$. We can easily verify that $g(x+y) = g(x) + g(y)$. But this is the Cauchy functional equation. Its only solution is $g(x) = ax$, from which the proof of the lemma follows. □

We have established that the functions $f_{01:11}$ and $f_{00:01}$ are affine but we can say more about their coefficients:

**Lemma 9.** *When $c_1 \neq 0$, there are constants $\lambda > 0$ and $\gamma_a$ such that*

$$f_{01:11} = \lambda t_{21} + \gamma_{01} - \gamma_{11},$$
$$f_{00:01} = \lambda t_{22} + \gamma_{00} - \gamma_{01},$$
$$f_{00:11} = \lambda(t_{21} + t_{22}) + \gamma_{00} - \gamma_{11}, \text{ for } c_1 > 0,$$
$$f_{01:10} = \lambda(t_{21} - t_{22}) + \gamma_{01} - \gamma_{10}, \text{ for } c_1 < 0.$$

*Moreover*
$\lambda = \frac{c_1}{c_2}$.

*Proof.* The three functions have the same multiplicative constant $\lambda$ because $f_{00:11}(t_{21}+t_{22}) = f_{01:11}(t_{21}) + f_{00:01}(t_{22})$ and all three functions are linear. It follows that $f_{01:11}(t_{21}) = \lambda t_{21} + \beta$ for some constant $\beta$. We can rewrite the constant $\beta$ as $\gamma_{01} - \gamma_{11}$. Similarly for the other functions.

The fact that $\lambda = \frac{c_1}{c_2}$ follows directly from the linearity and the semiperiodicity of the functions. For example, since $f_{01:11}(t_{21} + c_2) = f_{01:11}(t_{21}) + c_1$ and $f_{01:11}$ is linear it follows that $f_{01:11}(t_{21}) = \frac{c_1}{c_2} t_{21} + \beta$. □

From this and the fact that $c_1$ and $c_2$ have the same sign, we get:

**Lemma 10.** *When $c_1 \neq 0$, the payments of the first player (up to a common additive term which depends on $t_2$) are of the form $p_1(a_1, t_2) = -\lambda \cdot a_2 \cdot t_2 - \gamma_a$, for some constants $\lambda > 0$ and $\gamma_a$.*

With the above payments, the mechanism is the following affine minimizer:
$$\operatorname*{argmin}_{a} \{a_1 \cdot t_1 + \lambda \cdot a_2 \cdot t_2 + \gamma_a\}.$$

## 4 The case of many tasks

The generalization of the characterization to more than two tasks is almost straightforward. Fix a truthful mechanism. For two distinct tasks $j_1$ and $j_2$ we will write $j_1 \sim j_2$ when there are some values for the other $m - 2$ tasks such that the mechanism restricted to tasks $j_1$ and $j_2$ is an affine minimizer (i.e., with the associated constant $c_1 \neq 0$). It should be stressed that we require the mechanism restricted to these two tasks to be an affine minimizer for *some* values of the other tasks, not necessarily for all values, but we are going to see that the two are equivalent.

Our aim is to show that the relation $\sim$ is transitive; since it is clearly symmetric, it essentially partitions the tasks into equivalence classes with the exception that classes of size one are not affine minimizer but task-independent mechanisms. Assume that $j_1 \sim j_2$ and $j_2 \sim j_3$. That is, assume that when we fix some values of the other tasks, the mechanism for tasks $j_1$ and $j_2$ is an affine minimizer and when we fix some (not necessarily the same) values of the other tasks the mechanism for tasks $j_2$ and $j_3$ is also an affine minimizer, not necessarily with the consistent coefficients. Our aim is to show that the coefficients are consistent. We start with the case of two tasks and then we generalize.

**Lemma 11.** *When $j_1 \sim j_2$, the payments of player 1 satisfy the following for allocations $a$ and $b$ that agree on all other tasks (i.e., tasks not in $\{j_1, j_2\}$):*
$$p_a(t_2) - p_b(t_2) = \lambda_{j_1,j_2} \cdot (a - b)t_2 + \zeta_{a:b},$$
*where $\lambda_{j_1,j_2} > 0$ and $\zeta_{a:b}$ are constants.*

*Proof.* For each task $j$ not in $\{j_1, j_2\}$ we consider inputs with $t_{1j} = \infty$ or $t_{1j} = -\infty$ if $a_j = 0$ or $a_j = 1$, respectively. These inputs have allocations that agree with $a$ and $b$ on tasks not in $\{j_1, j_2\}$. For a fixed value then of the tasks not in $\{j_1, j_2\}$, the mechanism allocates tasks $j_1$ and $j_2$ with an affine minimizer. We therefore have
$$p_a(t_2) - p_b(t_2) = \lambda_{j_1,j_2} \cdot (a - b)t_2 + \zeta_{a:b},$$
where $\lambda_{j_1,j_2} > 0$ and $\zeta_{a:b}$ may depend only on the values of the other tasks.

The crucial observation is that when $a$ and $b$ differ in only one task these are constants (and do not depend on other tasks). It remains to show that this also holds when $a$ and $b$ differ on both tasks. However, if $a'$ is the allocation which differs from $a$ on task $j_1$, we have that
$$p_a(t_2) - p_{a'}'(t_2) = \lambda_{j_1,j_2} \cdot (a - a')t_2 + \zeta_{a:a'}$$
$$p_{a'}'(t_2) - p_b(t_2) = \lambda_{j_1,j_2} \cdot (a' - b)t_2 + \zeta_{a':b},$$

from which we get that $\zeta_{a:b} = \zeta_{a:a'} + \zeta_{a':b}$ is also constant since it is equal to the sum of two constants. □

We now generalize this lemma to many tasks.

**Lemma 12.** *When $j_1 \sim j_2$, $j_2 \sim j_3$, ..., $j_{k-1} \sim j_k$, the payments of player 1 satisfy the following for allocations $a$ and $b$ that agree on all other tasks (i.e., not in $\{j_1, \ldots, j_k\}$):*

$$p_a(t_2) - p_b(t_2) = \lambda_{j_1, \ldots, j_k} \cdot (a-b)t_2 + \zeta_{a:b},$$

*where $\lambda_{j_1, \ldots, j_k} > 0$ and $\zeta_{a:b}$ are constants.*

*Proof.* We show the lemma for $k = 3$ since the generalization is straightforward. We first show that the $\lambda$ coefficients are equal. We know that

$$p_a(t_2) - p_b(t_2) = \lambda_{j_1, j_2} \cdot (a-b)t_2 + \zeta_{a:b} \tag{1}$$

$$p_{\hat{a}}(t_2) - p_{\hat{b}}(t_2) = \lambda_{j_2, j_3} \cdot (\hat{a}-\hat{b})t_2 + \zeta_{\hat{a}:\hat{b}} \tag{2}$$

when $a$ and $b$ agree on all tasks not in $\{j_1, j_2\}$ and $\hat{a}$ and $\hat{b}$ agree on all tasks not in $\{j_2, j_3\}$. But the above sets of equations overlap when $a$ and $b$ differ only on task $j_2$. Therefore $\lambda_{j_1, j_2} = \lambda_{j_2, j_3}$ (we call this constant $\lambda_{j_1, j_2, j_3}$).

The proof of the lemma for the $\zeta$ terms, is identical to the proof of the previous lemma (the case of two tasks): Let $a'$ be the allocation which differs from $a$ in task $j_1$. With the same argument as in the previous proof, we conclude that $\zeta_{a:b} = \zeta_{a:a'} + \zeta_{a':b}$. This shows that $\zeta_{a:b}$ is constant. □

The relation $\sim$ is symmetric and transitive and it partitions the tasks into equivalence classes. Suppose for simplicity that all tasks belong to one class. Then the mechanism is an affine minimizer (when there are at least 2 tasks). This follows from the last lemma: Fix $b = 1$, i.e. in $b$ all tasks are allocated to player 1. The payment $p_b$ can be set arbitrarily, so we set it to some arbitrary constant $\gamma_b$. Then $p_a(t_2) = \lambda \cdot (a-b) \cdot t_2 + \zeta_{a:b} + p_b(t_2) = -\lambda \cdot a_2 \cdot t_2 - \gamma_a$, where we defined $\gamma_a = -\zeta_{a:b} + \gamma_b$ (a constant) and used $\lambda > 0$ as an abbreviation of $\lambda_{1,\ldots,m}$. Then the allocation for player 1 is given by

$$\underset{a_1}{\operatorname{argmin}}\{a_1 t_1 - p_a(t_2)\} = \underset{a_1}{\operatorname{argmin}}\{a_1 t_1 + \lambda a_2 t_2 + \gamma_a\},$$

with $\lambda$ and $\gamma_a$ constants.

The above lemma allows as to partition the tasks so that each part is independent of the other parts. Parts that have 2 or more tasks are affine minimizers. Parts that have only 1 task are not necessarily affine minimizers.

## 5 Lower bound for 2 tasks

Although our characterization involves only decisive mechanisms and negative values, it can be extended directly to show that the approximation ratio even for two tasks is at least 2. The following claim from [15] shows a non-decisive mechanism for positive values has unbounded ratio:

Suppose for example that the allocation 10 does not occur for some $t_2$, and take the input $\begin{pmatrix} \epsilon & \infty \\ t_{21} \star & t_{22} \star \end{pmatrix}$. Since the allocation of the first player cannot be 10 the allocation is indicated by the stars. By monotonicity the allocation is the same for the instance $\begin{pmatrix} \epsilon & \infty \\ t_{21} \star & \epsilon \star \end{pmatrix}$. But this gives approximation ratio $1 + t_{21}/\epsilon \to \infty$.

The following theorem reproduces the result in [15] for any number $m \geq 2$ of tasks.

**Theorem 3.** *No truthful mechanism for 2 players with $c_1 \neq 0$ can have a bounded approximation ratio. Consequently any mechanism for 2 players with bounded approximation ratio is a task independent mechanism.*

*Proof.* The reason is that for small values of $t_{21}$ and $t_{22}$, the constant $c_1$ dominates the algebraic expressions of the mechanism and as a result the mechanism is far from optimal.

Towards a contradiction suppose that a mechanism with $c_1 \neq 0$ has bounded approximation ratio $r$. We essentially look at the one dimensional case. Specifically, consider the case of $t_{12} = \epsilon$. In this way the optimal cost for the second task is almost zero and we can concentrate on the first task. The mechanism gives the first task to player 1 iff $t_{11} \leq f_{10:00}(t_{21}) + c_1$. For the instances with $t_{11} = f_{10:00}(t_{21}) + c_1 \pm \delta$, for some arbitrarily small $\delta$, the approximation ratio is at least $\frac{\max\{t_{21}, f_{10:00}(t_{21}) + c_1\}}{\min\{t_{21}, f_{10:00}(t_{21}) + c_1\}}$. So we must have

$$t_{21}/r \leq f_{10:00}(t_{21}) + c_1 \leq rt_{21}.$$

Similarly, if we consider the case $t_{22} = \epsilon$, we get that the first player gets the first task iff $t_{11} \leq f_{10:00}(t_{21})$, from which we can conclude that

$$t_{21}/r \leq f_{10:00}(t_{21}) \leq rt_{21}.$$

By subtracting the above inequalities and letting $t_{21}$ to tend to 0, it is clear that the above inequalities cannot hold unless $c_1 = 0$. □

We can now show that even for 2 tasks and 2 players, no mechanism can have approximation ratio less than 2.

**Theorem 4.** *No mechanism for 2 players and 2 tasks has approximation ratio less than 2.*

*Proof.* By Theorem 3 the mechanism is task-independent. Suppose that when the processing times of both players for the first task are both 1, player 1 gets it. Then the allocation for the instance $\begin{pmatrix} 1 \star & 1 \star \\ 1 & \infty \end{pmatrix}$ is indicated by the stars and the resulting approximation ratio is 2. (In the other case we take the matrix $\begin{pmatrix} 1 & \infty \\ 1 \star & 1 \star \end{pmatrix}$) □

In fact, for 2 tasks, we can show that the only truthful mechanism which achieves approximation ratio 2 is the VCG mechanism.

**Theorem 5.** *The only truthful mechanism for 2 players with approximation ratio 2 is the VCG mechanism.*

*Proof.* By symmetry, it suffices to show $f_{01:11}(t_{21}) = t_{21}$. Consider the instance

$$\begin{pmatrix} f_{01:11}(t_{21}) - \epsilon \star & t_{21} \star \\ t_{21} & \infty \end{pmatrix}.$$

The second task is allocated to player 1 (otherwise the approximation ratio of the mechanism is infinite) and the first tasks is also allocated to player 1 as the value of the first player does not exceed the threshold $f_{01:11}(t_{21})$. Therefore, by letting $\epsilon$ tend to 0, the approximation ratio is at least

$$\frac{f_{01:11}(t_{21}) + t_{21}}{t_{21}}.$$

This ratio is at most 2, only when $f_{01:11}(t_{21}) \leq t_{21}$. Consider also the instance

$$\begin{pmatrix} f_{01:11}(t_{21}) + \epsilon & \infty \\ t_{21} \star & f_{01:11}(t_{21}) \star \end{pmatrix}.$$

(This is similar to the previous instance, where we exchanged the 2 players). By letting $\epsilon$ tend to 0, the approximation ratio is at least

$$\frac{f_{01:11}(t_{21}) + t_{21}}{f_{01:11}(t_{21})}.$$

This ratio is at most 2, only when $f_{01:11}(t_{21}) \geq t_{21}$. In conclusion, the mechanism has approximation ratio at most 2 only when $f_{01:11}(t_{21}) = t_{21}$. □

## 6 Concluding remarks

We gave a characterization of decisive truthful mechanisms. What happens for non-decisive mechanisms? For two tasks we have the following cases:

**Mechanisms with some oblivious player** A mechanism where one player is decisive and the other is not, as in Example 3.

**Mechanisms decisive for only 3 allocations** Our proof can be extended to this case and shows that the only mechanisms are affine minimizers. An example is the mechanism with only three allocations: 00, 01, 11 and $f_{01:11}(t_{21}) = t_{21}$, $f_{00:01}(t_{22}) = t_{22}$, $f_{00:11}(t_{21} + t_{22}) = t_{21} + t_{22}$ and $c_1 = \infty$.

**Mechanisms with only 2 allocations** Consider the mechanism which gives either both tasks to player 1 or both tasks to player 2. It gives both tasks to player 1 iff $t_{11} + t_{12} \geq h(t_{21} + t_{22})$ for some increasing function $h$. This is a mechanism which is neither affine minimizer nor task independent when $h$ is not linear. (In this case we treat two tasks as a single task, so things are like in a single-parameter domain.)

Can our characterization be extended to more than two players? Threshold mechanisms seem to play a central role in the characterization of truthful mechanisms. In fact, we can show that they are exactly the additive mechanisms, i.e. mechanisms where the payment to player $i$ can be expressed as the sum of payments, one for every task allocated to player $i$. Notice that the definition of threshold mechanisms involves the allocation and the definition of additive mechanisms involves only the payments. We leave the proof of the next theorem for the full version of the paper.

**Theorem 6.** *A mechanism is additive iff it is a threshold mechanism.*